\documentclass[aps,twocolumn]{revtex4}

\usepackage{graphicx}
\usepackage[usenames]{color}
\usepackage{amsmath}

\usepackage{natbib}

\newcommand{\fref}[1]{Fig.~\ref{#1}}

\newcommand{\eref}[1]{Eq.~\ref{#1}}

\begin{document}

\title{Optical Spintronics in Organic-Inorganic Perovskite Photovoltaics}
\author{Junwen Li$^{1,2}$, Paul M. Haney$^1$}

\affiliation{1.  Center for Nanoscale Science and Technology, National Institute of Standards and Technology, Gaithersburg, MD 20899 \\
2.  Maryland NanoCenter, University of Maryland, College Park, MD 20742, USA }
\begin{abstract}
Organic-inorganic halide CH$_3$NH$_3$PbI$_3$ solar cells have attracted enormous attention in recent years due to their remarkable power conversion efficiency.  When inversion symmetry is broken, these materials should exhibit interesting spin-dependent properties as well, owing to their strong spin-orbit coupling.  In this work, we consider the spin-dependent optical response of CH$_3$NH$_3$PbI$_3$.  We first use density functional theory to compute the ballistic spin current generated by absorption of unpolarized light.  We then consider diffusive transport of photogenerated charge and spin for a thin CH$_3$NH$_3$PbI$_3$ layer with a passivated surface and an Ohmic, non-selective contact.  The spin density and spin current are evaluated by solving the drift-diffusion equations for a simplified 3-dimensional Rashba model of the electronic structure of the valence and conduction bands.  We provide analytic expressions for the photon flux required to induce measurable spin densities, and propose that these spin densities can provide useful information about the role of grain boundaries in the photovoltaic behavior of these materials.  We also discuss the prospects for measuring the optically generated spin current with the inverse spin Hall effect.
\end{abstract}

\maketitle

\section{Introduction}

Organic-inorganic halide CH$_3$NH$_3$PbI$_3$ solar cells have attracted enormous attention in recent years due to their remarkable power conversion efficiency, currently standing at 20.1 \%  \cite{Green_ppra_2015}.
Since being first employed as photovoltaic absorbers in 2009 with an efficiency of 3.8 \% \cite{Kojima_jacs_2009},  the pace of development of this material is unprecedented.  The high efficiency is attributed to the optimal bandgap ($\approx 1.5$ eV), high absorption coefficient, and efficient charge transport properties.  Long charge carrier diffusion lengths have been observed: over 1 $\mu$m for a solution-processed mixed perovskite CH$_3$NH$_3$PbI$_{3-x}$Cl$_x$ \cite{Stranks_science_2013, Xing_science_2013}, and over $10~{\rm \mu m}$ in samples with improved crystalline quality \cite{shi2015low}.  There are several open questions regarding the photovoltaic properties of these materials, such as the reason for the long carrier lifetime in highly disordered samples \cite{Xing_science_2013,yin2014unusual}, the role of chlorine doping in the device performance \cite{yu2014role,colella2014elusive}, and the origin of hysteresis in the current-voltage curve under illumination \cite{Gottesman_jpcl_2014,Stoumpos_ic_2013,beil}.

A rather different and, so far, less explored aspect of these perovskites is their potential use in spintronics applications  \cite{Kim_pnas_2014}.  Because of the heavy Pb atom, spin-orbit coupling plays an important role in the electronic structure \cite{Even_jpcl_2013,even2014dft}.  Additionally, first-principles calculations indicate that a non-centrosymmetric crystal structure which hosts ferroelectric order is stable \cite{Frost_nl_2014,zheng2015rashba,amat2014cation}.  Measurements have revealed ferroelectric domain formation \cite{kutes_jpcl_2014,kim_jpcl_2015}, although the presence of ferroelectricity in these materials is still under debate \cite{Xiao_naturemater_2015,beil}.  We note that even in bulk centrosymmetric materials, interfaces, surfaces, and planar defects such as grain boundaries locally break inversion symmetry.  The combination of strong spin-orbit coupling and breaking of inversion symmetry immediately leads to interesting spin-dependent properties.  Structural (bulk) inversion symmetry breaking leads to Rashba \cite{rashba1960properties} (Dresselhaus \cite{dresselhaus1955spin}) spin-orbit coupling; both cases result in a momentum-dependent effective magnetic field.  The Rashba model has been shown to describe the electronic structure of non-centrosymmetric CH$_3$NH$_3$PbI$_3$ near the bandgap \cite{Stroppa_nc_2014}.  Recent work utilizes these effects to propose an implementation of the the Datta-Das spin field-effect transistor with perovskite materials \cite{kepenekian2015rashba,datta1990electronic}.

Since CH$_3$NH$_3$PbI$_3$ exhibits both exceptional photovoltaic characteristics and useful spin properties, it should be an ideal material to study the optical excitation of spintronic effects.  In this work we consider optical excitation of spin currents and spin densities in CH$_3$NH$_3$PbI$_3$ by unpolarized light.  We consider the generation of ballistic pure spin current, which is associated with the non-equilibrium distribution of photo-excited electron-hole pairs before they undergo momentum relaxation.  The optical generation of ballistic spin current has been previously studied theoretically \cite{ganichev2003spin,tarasenko2005pure,Bhat_prl_2005}, and the induced spin densities have been measured in a variety of samples \cite{driel,Hubner_prl_2003,stevens2003quantum}.  We also consider the spin current and spin densities that result from the diffusion of photo-excited carriers.  The influence of spin-orbit coupling on diffusive carrier motion has been extensively studied in the context of 2-d Rashba systems \cite{burkov2004theory,halperin,gorini2010non}.  In this work we present a unified description of both the ballistic and diffusive spin-dependent response of CH$_3$NH$_3$PbI$_3$.  Our focus on this material is motivated by its exceptional properties, but we employ a generic Rashba model for our analysis so that the results apply to any three-dimensional Rashba semiconductor, such as GeTe \cite{di2013electric,chattopadhyay1987neutron} and BiTeI \cite{ishizaka2011giant}.

The spin-dependent responses of interest in this work naturally arise from a Rashba model.  An intuitive account of these can be formulated based on the spin texture of the conduction bands, as shown in \fref{fig:wannier_bands}(c).  States with momentum $+k_x$ and $-k_x$ have spins aligned in opposite directions.  If both states are occupied, there is a vanishing spin density, but a nonzero spin current (since the spin current is the product of spin and velocity).  A photoexcited distribution of electrons in the conduction band therefore leads to a nonequilibrium spin current.  If on the other hand there is an imbalance in the occupation of states at $+k_x$ and $-k_x$ (which corresponds to a current-carrying distribution), there is also an imbalance in spin, resulting in a spin accumulation in the negative $y$-direction (the Edelstein effect \cite{Edelstein_ssc_1990}).  The contribution of spin from the outer and inner Fermi circles partially cancel, but a nonzero net spin remains which is proportional to the charge current.  The results we derive can be understood roughly in terms of this simple picture.

The paper is organized as follows: In Sec.~\ref{sec:ballistic} we compute the ballistic spin current generation using density functional theory.  We find that the steady state optically excited ballistic spin current is proportional to the Rashba spin-orbit parameter, the optical absorption coefficient, and the momentum scattering time.  In Sec.~\ref{sec:diffusive} we solve the spin-dependent diffusion equations for a thin film CH$_3$NH$_3$PbI$_3$ layer with a passivated surface and an ideal front contact.  We provide analytic expressions for the spin current and spin density, demonstrating how these quantities scale with material parameters.  Finally, in Sec.~\ref{sec:discussion} we discuss the magnitude of the optically excited spin densities and spin currents, and provide estimates for the photon flux required to induce measurable spin densities and spin currents.  We discuss the possible uses of spintronic effects in these materials, including using the optically induced spin density to probe the role of grain boundaries in the photovoltaic behavior of these materials.

\section{Electronic Structure and Ballistic Spin Current}  \label{sec:ballistic}

In this section we present first-principles results of the ballistic spin current generated by optical absorption in CH$_3$NH$_3$PbI$_3$.  We begin by discussing the electronic structure of this material, and demonstrating the applicability of the Rashba model for the conduction and valence bands.  We then evaluate the spin current generation rate response function, and compare detailed first-principles results with a simple estimate of this response function provided by the Rashba model.

\begin{figure}[h]
  \includegraphics[angle=-90,scale=0.31]{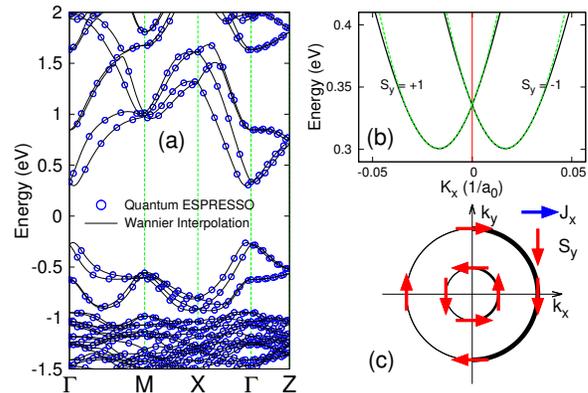}
  \caption{(a) Comparison between Wannier interpolated (solid line) and Quantum-ESPRESSO (dots) bands.  (b) Zoom in of the conduction bands near the $\Gamma$ point, along with fitting to the Rashba Hamiltonian.  The fitting parameters are $m^*_c=0.11~m_e$, $\alpha_c=-0.22~{\rm eV\cdot nm}$.  For the valence band (not shown), the fitting parameters are:  $m^*_v=0.14~m_e$, $\alpha_v=0.15~{\rm eV\cdot nm}$. (c) Cartoon of constant energy circles for $k_z=0$, showing the ${\bf k}$-dependent spin texture of the conduction bands.  A current-carrying distribution results in more states occupied in the $k_x>0$ half of the plane (denoted by a thicker line), and a net spin accumulation in the -$y$-direction.}
  \label{fig:wannier_bands}
\end{figure}

We perform first-principles density functional theory calculations using local density approximation (LDA) in the form of norm-conserving pseudopotentials as implemented in Quantum-ESPRESSO \cite{Paolo_jpcm_2009}.  We use an energy cutoff of 80 Ry for the plane wave basis expansion; for the structural relaxation, a $4 \times 4 \times 3$ grid for the Brillouin zone sampling was employed; all atoms in the unit cell were allowed to move until the force on each is less than $0.5~{\rm eV/nm}$.  The lattice constants are calculated to be $a = 0.875 ~{\rm nm}$ and $c = 1.203~{\rm nm}$, in good agreement with the experimental measurements ($a = 0.880~{\rm nm}$  and $c = 1.269~{\rm nm}$) \cite{Kawamura_jpsj_2002}. We investigate the tetragonal phase of space group $I4cm$.  The Pb atoms are displaced along the $+\bf z$-direction relative to the octahedral center.  This lattice structure is non-centrosymmetric and exhibits ferroelectricity.  Using the Berry phase approach, we find an electrical polarization of 10.7~$\mu$C$/$cm$^2$. The band structure along high symmetry directions is shown in \fref{fig:wannier_bands}(a).  To test the influence of the ferroelectric order on the results, we've also studied a crystal with a strongly reduced ferroelectric polarization of $0.31~{\rm \mu C/cm^2}$.

The combination of strong spin-orbit coupling (derived from the Pb atom) together with bulk inversion asymmetry leads to conduction and valence band states which are well described by a 3-dimensional Rashba model \cite{Kim_pnas_2014}.  The effective Hamiltonian is:
\begin{eqnarray}
H^{\rm eff}_{c,v}=\frac{\hbar^2 {\bf k}^2}{2 m^*_{c,v}} + \alpha_{c,v}\left[ {\bf \sigma} \cdot \left({\bf k} \times {\bf z}\right)\right],\label{eq:rashbaH}
\end{eqnarray}
where ${\bf z}$ is the broken symmetry direction, ${\bf k}$ is the three-dimensional Bloch wave vector, ${\bf \sigma}$ is the vector of Pauli spin matrices, $m^*$ is the effective mass, $\alpha$ parameterizes the Rashba spin-orbit coupling, and the $c,v$ subscripts correspond to conduction and valence band.  The Rashba parameter $\alpha_{c,v}$ depends on the degree of inversion symmetry breaking, and vanishes for inversion symmetric systems.  The Rashba spin-orbit term acts as a ${\bf k}$-dependent effective magnetic field along the direction perpendicular to ${\bf k}$ and ${\bf z}$.  \fref{fig:wannier_bands}(b) shows the fitting of the conduction band structure to the Rashba Hamiltonian.  We'll refer to this effective model to gain an intuitive understanding of the ballistic spin current magnitude, and as a starting point for the description of charge and spin diffusion.  All results are presented in terms of $\alpha_{c,v}$ and can therefore be applied to cases where inversion symmetry is more weakly broken (for example, in cubic systems with aligned CH$_3$NH$_3$ dipoles \cite{leijtens2015modulating,frost2014molecular}, or in bulk inversion symmetric systems near interfaces).

We next compute the optically generated spin current.  We consider the response of the system to a monochromatic electric field of frequency $\omega$
\begin{equation}
\mathbf{E}(t) = \mathbf{E}(\omega) e^{-i\omega t} + \mathbf{E}^*(\omega) e^{i\omega t}
\end{equation}
and employ the symmetrized spin current operator defined as $\hat{Q}^{ij} = \frac{1}{2}\{\hat{v}^i, \hat{s}^j\}$ \cite{Zhou_prb_2007,Rioux_prb_2014}, where $\hat{v}^i$ is the $i$-component of the velocity operator, and $\hat{s}^j$ is the $j$-component of the spin operator.  The spin current generation rate is derived by solving semiconductor optical Bloch equations perturbatively to first order in the field intensity \cite{Schafer_springer_2002}, and is given by:

\begin{equation}
\dot{Q}^{ij} = \mu^{ijlm}(\omega) E^{l*}(\omega) E^{m}(\omega) \rm, \label{eq:qdot}
\end{equation}
where
\begin{equation}
\begin{aligned}
\mu^{ijlm}(\omega) & = \frac{2\pi e^2}{\hbar^2 \omega^2} \frac{1}{V} \sum_{\mathbf{k}} \sum_{cv} ( Q^{ij}_c - Q^{ij}_v )   \\
& \times v^{l*}_{cv} (\mathbf{k}) v^{m}_{cv}(\mathbf{k})
 \delta[ \omega_{cv}(\mathbf{k}) - \omega ] \label{eq:mu1}
\end{aligned}
\end{equation}
In \eref{eq:mu1}, $v^{m}_{cv}\left({\bf k}\right)=\langle \psi_c\left({\bf k}\right) | \hat{v}^m | \psi_v\left({\bf k}\right) \rangle$ is the velocity operator matrix element  between conduction and valence band states, $Q^{ij}_{c(v)}=\langle \psi_{c(v)}\left({\bf k}\right) | \hat{Q}^{ij} | \psi_{c(v)}\left({\bf k}\right) \rangle$, and $\omega_{cv}({\bf k})= \left(E_c\left({\bf k}\right)-E_v\left({\bf k}\right)\right)/\hbar$.  Roman superscripts indicate Cartesian components and summation over repeated indices is implied.  To evaluate \eref{eq:mu1}, we employed Wannier-function techniques to calculate the band structure and momentum matrix elements on a very fine grid in momentum space \cite{Giustino_prb_2007,Yates_prb_2007,Marzari_rmp_2012}.  \fref{fig:wannier_bands}(a) shows the matching of the plane wave and Wannier function band structures.  We interpolate \eref{eq:mu1} to evaluate the pure spin current on a fine grid of $100 \times 100 \times 75$ $k$-points. The accuracy of this method has been verified on a coarse grid by comparing the pure spin current responses calculated with the Wannier interpolation technique and plane wave basis.

\begin{figure}[!h]
  \includegraphics[angle=-90,scale=0.3]{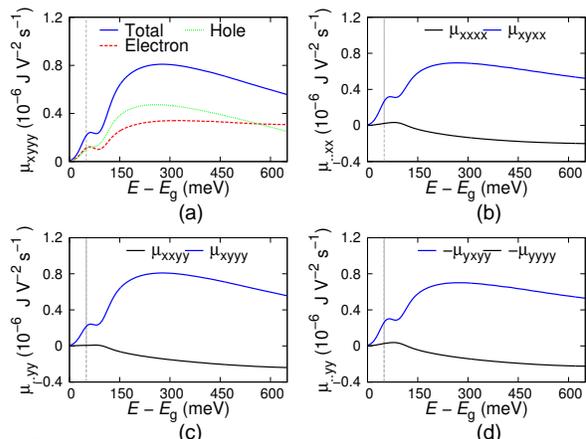}
  \caption{(a) The total spin current response versus photon energy (subtracted by the bandgap energy $E_g$), along with the separate electron and hole contributions (b) the total spin current response for electron velocity in the $x$-direction, and light polarized in the $x$-direction.  (c) the same as (b), but with light polarized in the $y$-direction.  (d) the total spin current response for electron velocity in the $y$-direction, and light polarized in the $y$-direction.}
  \label{fig:mu1222}
\end{figure}

In \fref{fig:mu1222} we show the $\mu^{xyyy}$ component of spin current response pseudotensor, corresponding to the spin current moving along the $x$-direction with spin oriented along the $y$-direction under the illumination of light linearly polarized along $y$-direction.  We find that the spin current response of electrons and holes has the same sign.  This is the result of the opposing sign of the effective Rashba parameter in the valence and conduction bands.  We note that other works have identified the significance of the relative sign of the conduction and valence band Rashba parameters to the optical response \cite{zheng2015rashba}.  Figs.~\ref{fig:mu1222}(b) and (d) show that there is nearly an identical spin current response for light polarized along the $x$ and $y$-directions.  This is due to the approximate equivalence of $x$ and $y$-directions in the crystal (recall $z$ is the symmetry-breaking direction). This shows that {\it unpolarized} light can effectively generate ballistic spin current in this material.

An order of magnitude estimate of $\mu$ is provided by noticing the similarity of \eref{eq:mu1} to the imaginary part of the dielectric response.  Considering linearly polarized light (along the $y$-direction), and taking the spin current expectation value to be independent of ${\bf k}$, ($Q^{ij}_{c(v)}\left(\bf k\right)=\bar{Q}^{ij}$), \eref{eq:qdot} can be written
\begin{eqnarray}
\dot{Q}^{ij}\left(\omega\right)\approx\bar Q^{ij} \beta\left(\omega\right) \phi_0 \approx  \left(\alpha_c-\alpha_v\right) \beta\left(\omega\right) \phi_0 \label{eq:qdotRashba}
\end{eqnarray}
where $\beta\left(\omega\right)$ is the absorption coefficient, and $\phi_0$ is the incident photon flux ($\phi_0=\epsilon c E^2 / \left(2 \hbar \omega\right)$).  The second approximate relation shown in \eref{eq:qdotRashba} follows from evaluating the spin current in the Rashba model, which scales as $\alpha$.  From the ground state electronic structure, we find $\alpha_c-\alpha_v = -0.37~{\rm eV\cdot nm}$ and $\beta=\left(230~{\rm nm}\right)^{-1}$, leading to an estimated spin current response $\mu$ value of $-8.4\times 10^{-6}~{\rm J/\left(V^2\cdot s\right)}$.  The first-principles result has a different sign, which is traced back to larger dipole transition matrix elements for states in the inner Fermi circle relative to the outer Fermi circle (see Fig. \ref{fig:wannier_bands}(c)).  The magnitude of the simple estimate is about 10 times smaller found in the first-principles calculation (see \fref{fig:mu1222}).  The discrepancy in magnitude is mostly attributed to the approximation for the spin current value, but \eref{eq:qdotRashba} provides a crude estimate.

\begin{table*}
\begin{center}
\begin{tabular}{|l| c | c | c| }
\hline
&  ~~ Purely Ballistic~~   & Purely Diffusive  &  Diff. with only $\nabla Q_{\rm ball}$ source  \\
   & ~$\left(\Gamma_{sc}=0,{\tilde \mu}\neq 0,\beta L \ll 1 \right)$   & $\left(\Gamma_{sc}\neq 0,{\tilde \mu}=0,\beta L \ll 1 \right)$ & ~~~$\left(\Gamma_{sc}=0,{\tilde \mu}\neq 0,\beta L > 1\right)$~~~ \\ \hline

 ~~Spin current:~ $\begin{aligned}[c] \frac{Q^{yx}}{ \phi_0}\end{aligned}$
 &  $\begin{aligned}[c]\frac{\alpha\beta\tau_k}{\hbar}~(\approx 9\times10^{-4})\end{aligned}$
 & $\begin{aligned}[c]~~\beta\Gamma_{sc}\sqrt{\frac{\tau_s}{D}}L_d \tanh\left(\frac{L}{L_d}\right)~(\approx0.04)\end{aligned}$  ~~
 & $\begin{aligned}[c]\frac{\alpha \beta^2 \tau_k \sqrt{D\tau_s}}{\hbar} ~(\approx4.4\times 10^{-5})\end{aligned}$\\ \hline

  ~~Spin Density:~$\begin{aligned}[c]\frac{S_x}{ \phi_0\beta\tau_n}\end{aligned}$~
  & 0
  & $\begin{aligned}[c]\frac{\Gamma_{sc}\tau_s}{L_d}~(\approx2.6\times 10^{-4})\end{aligned}$
  &  $\begin{aligned}[c]\frac{\alpha\beta\tau_s \tau_k}{\hbar \tau_n}~(\approx2.3\times 10^{-7})\end{aligned}$ \\ \hline

\end{tabular}
\end{center}
\caption{Scaling of the photogenerated spin current and spin density which arises from: the optically generated ballistic spin current (first column), the diffusion of optically generated carriers (second column), and the diffusion of spins created from ballistic spin current gradient (third column).  The spin current is scaled by the photon flux, while the spin density is scaled by the generation rate density times the carrier lifetime.  The Rashba parameter $\alpha$ should be replaced with $\alpha_{c(v)}$ to evaluate the electron (hole) contribution.  The approximate value of each expression (for electron plus hole contributions) is denoted in parenthesis, for material parameter values of: $\tau_k=1.5\times 10^{-15}~{\rm s}$, $\alpha_c=-0.02~{\rm eV\cdot nm}$, $\alpha_v=0.06~{\rm eV\cdot nm}$, $\tau_n=10~{\rm ns}$, $D=0.625~{\rm cm^2/s}$, $\beta=\left(230~{\rm nm}\right)^{-1}$.  Derived parameter values are $\Gamma_{sc}^e=8~{\rm m/s}$, $\Gamma_{sc}^h=-152~{\rm m/s}$, $\tau_s^e=7\times10^{-12}~{\rm s}$, $\tau_s^h=10^{-12}~{\rm s}$, $L_d=790~{\rm nm}$.  Recall $\Gamma_{sc}\propto \alpha^3\tau_k^2$, $\tau_s \propto \left(\alpha^2 \tau_k\right)^{-1}$.  Rashba parameters are chosen such that the resulting spin lifetimes match those measured in Ref. \cite{Giovanni_nl_2015}}
\label{tab_1}
\end{table*}

We next discuss the role of ferroelectric order in the spin current response function.  To study this, we calculate the spin current response for a lattice structure with a computed ferroelectric polarization of $0.31~{\rm \mu C/cm^2}$, or about 30 times smaller than the previous case.  The primary difference between the two crystal structures is that in the previous, highly polarized case, the CH$_3$NH$_3$ ions are mostly aligned in the $z$-direction, whereas in the weakly polarized crystal, the ions are randomly oriented.  For the weakly polarized material, the maximum spin current response is $3.4\times 10^{-7}~{\rm J/\left(V^2\cdot s\right)}$, a bit less than half of the value found for the highly polarized material (see \fref{fig:mu1222}(a)).  This shows that the spin current response is not proportional to the ferroelectric polarization.  On the other hand, the value of $\alpha_c - \alpha_v$ for the weakly polarized material is $-0.19~{\rm eV\cdot nm}$, approximately half that of the highly polarized material.  Thus the spin current response scales with the magnitude of the effective Rashba parameter, further validating of \eref{eq:qdotRashba}.  In summary, the spin current response is determined by $\alpha$, and $\alpha$ is determined by the degree of inversion symmetry breaking, which is not necessarily correlated with the ferroelectric polarization value.

As described in Ref.~\onlinecite{Hubner_prl_2003}, momentum scattering destroys the ballistic current generation computed with \eref{eq:mu1}.  This is due to the misalignment of the spin and the effective magnetic field after a momentum-changing scattering event.  The misalignment results in the spin {\it precessing} around the new effective magnetic field, and after several scatterings, there is a partial decoupling of electron momentum and spin.  Here we consider the steady state ballistic spin current, which is given by $Q^{ij}=\dot{Q}^{ij} \tau_k$, where $\tau_k$ is the momentum relaxation time.  We assume that $\tau_k$ is shorter than all other time scales (generally we take $\tau_k\approx 1~{\rm fs}$), so the ballistic current is equal to its steady state value while the sample is under illumination, and vanishes otherwise.  In Table~\ref{tab_1}, we show the scaling of the ballistic current (scaled by $\phi_0$) based on the Rashba approximation described in the previous paragraph.  Based on material parameters of CH$_3$NH$_3$PbI$_3$ given in the caption of Table~\ref{tab_1}, we estimate a spin current conversion ``efficiency'' (or spin flux per photon flux) of approximately $9\times10^{-4}$.  In Sec. \ref{sec:discussion}, we discuss the magnitude of this spin current in terms of measurements and spintronics applications.

\section{diffusive spin current} \label{sec:diffusive}

We next consider the diffusive spin response to optical excitation, which applies for time scales longer than the momentum relaxation time.  We utilize the model developed in Ref.~\onlinecite{burkov2004theory}, which describes the diffusion of carriers subject to the Rashba Hamiltonian of \eref{eq:rashbaH}.  As carriers diffuse from one impurity to another in momentum-changing scattering events, their spin precesses around the instantaneous effective magnetic field (which is perpendicular to $\bf z$ and the momentum $\bf k$).  A carrier's real space random walk therefore leads to a random walk in spin space, and ultimately results in Dyakonov-Perel spin relaxation \cite{Dyakonov_spss_1972}.  Additionally, the Rashba interaction leads to spin-charge coupling, whereby a diffusive charge current induces spin accumulation - a diffusive version of the Edelstein effect \cite{Edelstein_ssc_1990}.  Gradients in this spin accumulation subsequently result in diffusive spin current, which is distinct from the ballistic spin current calculated in the previous section.

We consider the limit of $k_F\tau_k\alpha/\hbar \ll 1$.  This corresponds to a small spin precession between scattering events.  As discussed in Ref.~\onlinecite{burkov2004theory}, for diffusion along the $y$-direction, the charge and $x$-component of spin decouple from the $y$ and $z$-components of spin. We therefore restrict our attention to the charge and $x$-component of spin $S_x$.  An incident monochromatic photon flux $\phi_0$ and absorption coefficient $\beta$ lead to a position-dependent electron-hole pair generation rate density given by $\phi_0 \beta \exp\left(-\beta y\right)$.  The steady state diffusion equations for photogenerated electron number density $n$ and number density of spin $S_x$ are then given as:
\begin{eqnarray}
-D\frac{\partial^2 n}{\partial y^2} + \frac{n}{\tau_n} + \Gamma_{sc} \frac{\partial S_x}{\partial y}  &=& \phi_0\beta  \exp\left(-\beta y\right)\label{eq:dd1} \\
-D\frac{\partial^2 S_x}{\partial y^2} + \frac{S_x}{\tau_s} + \Gamma_{sc} \frac{\partial n}{\partial y}  &=&  \frac{\partial Q^{yx}_{\rm ball}}{\partial y}\label{eq:dd2}
\end{eqnarray}
where $\tau_n$ is the carrier lifetime (or electron-hole pair recombination time), $D$ is the diffusivity, $\tau_s$ is the spin lifetime, and $\Gamma_{sc}$ parameterizes the charge-spin conversion, which results from the Rashba interaction.  In terms of microscopic parameters, $D=v_F^2 \tau_k/3$, $\tau_s=3\hbar^2/\left(2\alpha^2 k_F^2 \tau_k\right)$, and $\Gamma_{sc}=-\left(7\pi/32\right)\alpha^3 k_F^2 \tau_k^2 / \hbar^3$, where $v_F$ and $k_F$ are the Fermi velocity and wave vector, respectively.  Note that the numerical prefactors in these expressions differ from those in Ref.~\onlinecite{burkov2004theory} due to the system dimensionality and free electron dispersion along the $z$-direction \cite{footnote}.

The source term on the right hand side of \eref{eq:dd2} is the divergence of the ballistic spin current we evaluated in the previous section.  We've added this term ``by hand", motivated by the general form of a continuity equation.  This procedure is not generally valid because spin-orbit coupling leads to non-conservation of spin.  Including the source term in this way amounts to assuming that the divergence of the ballistic spin current is transferred entirely to spin density, and none of its angular momentum is transferred to the lattice \cite{haney2010current}.  Our analysis therefore represents an upper bound on the conversion of the ballistic spin current to spin accumulation.  We use the estimate of the ballistic spin current provided in the previous section: $Q_{\rm ball}^{yx}=\alpha\beta\tau_k \phi_0 \tilde{\mu} \exp\left(-\beta y\right)$, where $\tilde{\mu}$ is a dimensionless parameter we introduce to artificially tune the strength of the ballistic spin current.

\begin{figure}[!h]
  \includegraphics[scale=0.55]{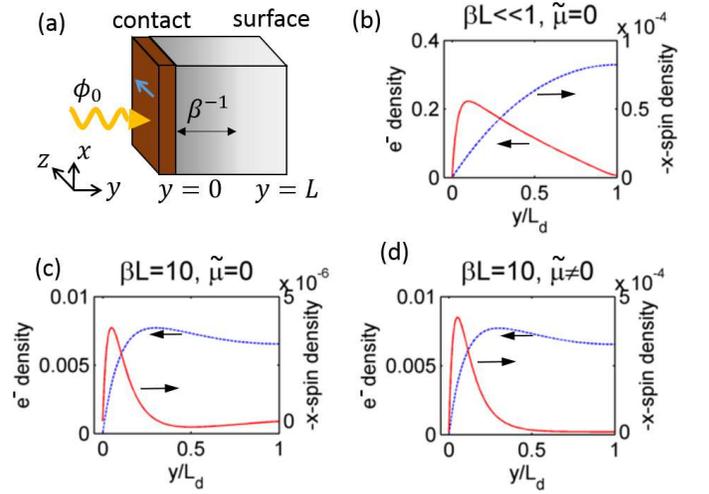}
  \caption{(a) shows the system geometry.  The gradient in shading of the sample indicates the length scale of generation electron-hole pair generation, $\beta^{-1}$.  The blue arrow on the contact in the $z$-direction denotes the direction of symmetry breaking in the perovskite, and the direction of the current induced by the inverse spin Hall effect.  (b) is the charge and spin density for $\beta L\ll 1$, $\tilde{\mu}=0$.  Densities are scaled by $\phi_0 \beta \tau_n$.  (c) shows the same for $\beta L=10$.  (d) shows the charge and spin density for an unrealistically large value of $\tilde{\mu}$ (100 times larger than the expected value), as an illustration of the effect of a spatially varying ballistic spin current on the diffusion.}
  \label{fig:dd}
\end{figure}

The system geometry is shown in \fref{fig:dd}(a).  Diffusion occurs along the $y$-direction, and we assume a non-selective, Ohmic contact at $y=0$, and a perfectly passivated surface at $y=L$.  The bulk symmetry breaking direction of the material is the $z$-direction.

As discussed in the previous section, both conduction and valence bands are well described by the Rashba model.  We present the analysis for the electrons here; the behavior of the holes is similar, and is also described by equations of the form of Eqs. \ref{eq:dd1}-\ref{eq:dd2}.  The relative sign for the Rashba parameter for electrons and holes plays a key role in determining the spin accumulation.  In this case, since electrons and holes have opposite signs of the Rashba parameter, electrons and holes diffusing in the same direction lead to a spin accumulation in the same direction.

The boundary conditions and general solution to Eqs. \ref{eq:dd1}-\ref{eq:dd2} are given in the appendix.  Analytical expressions are available in the limit of $\beta L \ll 1$ (uniform generation), $\tau_s \ll \tau_n$, and vanishing ballistic spin current ($\tilde{\mu}=0$).  The diffusive electron and spin number currents collected at $y=0$ take the form:
\begin{eqnarray}
\frac{J_c}{\phi_0} &=& \beta L_d \tanh\left(\frac{L}{L_d}\right) \\
\frac{Q^{yx}_{\rm diff}}{\phi_0}&=& \frac{\Gamma_{sc}}{\sqrt{D/\tau_s}}~ \beta L_d \tanh\left(\frac{L}{L_d}\right)
\end{eqnarray}
where $L_d=\sqrt{D\tau_n}$ is the diffusion length.

The electron carrier and spin densities for this case are shown in \fref{fig:dd}(b) (both carrier and spin are given in number densities, scaled by $\phi_0 \beta \tau_n$).  For the spatially uniform generation rate density considered here, the diffusive electron current varies linearly throughout the device thickness.  This results in a spin accumulation which increases linearly from the back contact.  This spin accumulation is a manifestation of the diffusive Edelstein effect (note that the spin is oriented along the $-\hat x$-direction).  The collecting contact at $y=0$ absorbs the nonequilibrium spin. The sharp gradient of the spin density near $y=0$ determines the collected spin current; the length scale of this steep drop is the spin diffusion length $\sqrt{D \tau_s}$ ($\approx 2~{\rm nm}$).  The scaling of the collected diffusive spin current and nonequilibrium spin density is shown in the second column of Table~\ref{tab_1}.  For the parameters given in the caption, we find a much larger conversion efficiency for the diffusive spin current compared to the ballistic spin current.  (We emphasize that the magnitude of the diffusive spin current depends on the geometry and boundary conditions.)

We next consider a non-uniform generation rate density with $\beta L=10$ and a vanishing ballistic spin current generation ($\tilde{\mu}=0$).  The result shown in \fref{fig:dd}(c) shows that, as expected, the charge current is developed near the front contact where the absorption occurs.  The rapid variation in $n$ near the contact results in a more rapid increase of the nonequilibrium spin.  The ratio of the spin current to electron current is nevertheless similar to the uniform generation case.

Finally, we add the ballistic spin current generation, $\tilde{\mu}\neq 0$.  For our CH$_3$NH$_3$PbI$_3$ parameters, we find the inclusion of the ballistic spin current has a negligible impact on the diffusive system response.  This is shown in the third column of Table~\ref{tab_1}.  The expressions for spin current and spin density given in the table are derived from solving \eref{eq:dd2}, with $\Gamma_{sc}=0$ and $\tilde{\mu}\neq0$.  They therefore capture only the contribution of the $\partial Q^{yx}_{\rm ball}/\partial y$ source term to the diffusive spin behavior.  It's shown that this contribution is quite small compared to the purely diffusive contributions.  In \fref{fig:dd}(d) we artificially increase the ballistic spin current magnitude by a factor of 100 over its estimated value in order to observe its effect.  We find that there's still negligible effect on the charge density, while the spin density and spin current increase in magnitude by about a factor of 100.

\section{Discussion} \label{sec:discussion}

An obvious question is whether the spin currents and densities computed in the last sections are significant enough to be measured, or even used for practical applications.  We first consider the optical measurement of the non-equilibrium spin density as a means to probe the diffusive charge current distribution, and discuss the usefulness of this information for polycrystalline photovoltaics.  We next consider measuring the optically generated spin current via the inverse spin Hall effect of the contact.  In both cases, a primary challenge is to induce sufficiently large spin responses.  We provide an estimate of the required photon fluxes in terms of microscopic material parameters.

Before proceeding we first comment on the magnitude of the Rashba parameters obtained from density functional theory.  The diffusive spin treatment of the last section relates the Dyakonov-Perel spin relaxation time to the Rashba parameters.  The Rashba parameters from density functional theory result in a spin relaxation time for electrons (holes) which is 90 (6) times smaller than that measured in Ref. \cite{Giovanni_nl_2015}, indicating that these Rashba parameter are unrealistically large.  Reducing $\alpha_c$ and $\alpha_h$ by a factor of approximately 9.4 and 2.4, respectively (corresponding to $\alpha_c=-0.02~{\rm eV\cdot nm},~\alpha_v=0.06~{\rm eV\cdot nm}$) leads to spin lifetimes that agree with experiment.  In the discussion below, we present numerical estimates related to the observation of spintronic effects using these reduced values of Rashba parameters.

\subsection{Spin densities}
We first consider the magnitude of the induced spin density.  Kerr measurements on the nonequilibrium spin density induced by the spin Hall effect in GaAs are able to detect spin densities on the order of $10^{13}~\hbar~{\rm cm^{-3}}$ \cite{stern,footnote2}.  The scaling of the spin density induced by diffusion in a 3-dimensional Rashba material is given in the second row of Table~\ref{tab_1}.  For the parameters given in the caption, a measurable diffusive spin density is attained for a photon flux $\phi_0\approx 9\times 10^{23}~{\rm m^{-2}~s^{-1}}$.  This flux is approximately 400 times greater than the incident solar flux of photons with energy greater than 1.5 eV (the bandgap of CH$_3$NH$_3$PbI$_3$).  The optically generated spin density is proportional to $\nabla n$; therefore measuring the spatial distribution of the spin accumulation enables the determination of the distribution of charge currents in a photovoltaic device.  We discuss the significance of this in the following.

\begin{figure}[!h]
  \includegraphics[angle=0,scale=0.6]{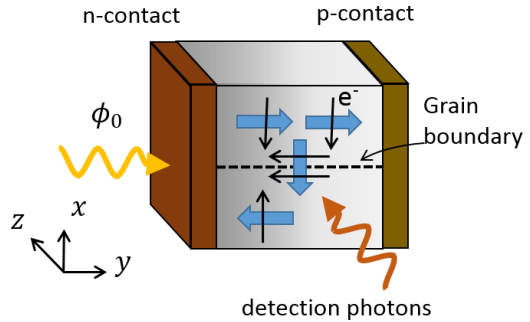}
  \caption{Depiction of system geometry used for the detection of spatially nonuniform charge diffusion due to the presence of grain boundaries (shown as dashed line). The perovskite absorber is taken to be p-type and sandwiched between selective contacts.  An electrostatic potential attracts minority carriers (electrons) to the grain boundary core, where they diffuse to the electron collecting contact.  In the schematic, the electron velocity is denoted with straight black arrows, and the resulting spin density is denoted with thick blue arrows.  The detecting photons are used to measure the spin density on the $x-y$ plane with the magneto-optical Kerr effect, and the crystal symmetry breaking direction of the grains must both be in the $z$-direction.  $\phi_0$ represents the photon flux which excites electron-hole pairs in the absorber.}
  \label{fig:GB}
\end{figure}

Many perovskite photovoltaics are polycrystalline; the material is permeated by grain boundaries, which are generally detrimental to charge carrier collection and photovoltaic performance due to their high defect density.  (In this discussion we assume the grain boundaries are oriented perpendicular to the charge collecting contacts, see Fig. \ref{fig:GB}.)  The high carrier collection which persists in spite of grain boundaries raises the prospect that grain boundaries could in fact play a beneficial role in photovoltaic performance \cite{yun2015benefit,li2015microscopic}.  This echoes similar scenarios put forth to explain the high photovoltaic conversion efficiency of other polycrystalline materials such as CdTe and copper indium gallium selenide (CIGS) \cite{jiang2004does,visoly2006understanding,azulay2007current,Persson_prl_2003,Yan_prl_2006,Li_prl_2014}.  Although the overall role of grain boundaries in these materials remains an open question \cite{taretto2008numerical,li2014grain,yun2015benefit}, it's generally accepted that charged grain boundaries which induce band bending sufficient to cause type inversion at the grain boundary core act as an efficient charge collectors rather than as recombination centers \cite{li2014grain} (under short circuit conditions).  In this case, the electrostatic field near a grain boundary separates electrons and holes, and the carrier which is attracted to the defective grain boundary core avoids recombination because of the type inversion which occurs there.  There is substantial minority carrier diffusion {\it toward} grain boundaries, and {\it along} the grain boundary cores towards the contacts \cite{footnote3}.  The diffusion induces spin accumulation with spin direction perpendicular to the diffusion current direction.  Obtaining the spatial distribution of charge diffusion via the spin density offers a route to testing this generally accepted picture of efficient polycrystalline photovoltaics.  We emphasize that this proposal requires the grains' crystal symmetry breaking direction is perpendicular to the sample surface (see Fig. \ref{fig:GB}).

\subsection{Spin currents}
We next consider the magnitude of the optically generated spin current.  Spin currents are measured only indirectly.  One way to detect spin currents is to observe their influence on ferromagnetic dynamics.  A dramatic example is current-induced magnetic switching: a sufficiently large flux of spin current into a thin ferromagnetic layer induces irreversible switching between easy axis orientations through an effect known as spin transfer torque \cite{slonczewski1996current,berger1996emission,ralph2008spin}.  To induce magnetic switching, the spin flux must exceed the magnetic damping rate per area.  For typical ferromagnet thin films, the required spin flux ($\approx 10^{29}~ \hbar~{\rm m^{-2}~s^{-1}}$ for a 1 nm film thickness) likely exceeds that which can be obtained through optical excitation of spin current.  This is due to higher order recombination processes (such as radiative and Auger recombination) which occur for large photogenerated carrier densities (typically greater than $10^{18}~{\rm cm^{-3}}$ \cite{strauss}).  These processes dramatically reduce carrier lifetimes and suppress the net generation rate for the large photon fluxes needed for magnetic switching.  In colloquial terms, current-induced magnetic switching can push metals to their conduction limits, and is therefore likely untenable for optically excited semiconductors.

An alternative route to detecting spin currents is through the inverse spin Hall effect.  In this case a spin current with real space motion along the $\hat v$-direction and spin along the $\hat s$-direction induces a charge current along the $\hat{v}\times\hat{s}$ direction.  This effect is parameterized by the spin Hall angle $\theta_{sh}$: $J_c = \theta_{sh} J_s$ (as before, we take $J_c$ and $J_s$ to be number current densities).  Heavy metals like Pt exhibit relatively large spin Hall values, on the order of 0.05 \cite{liu2011spin,mosendz2010detection,rojas2014spin}.  A nonzero spin Hall angle in the collecting contact of \fref{fig:dd}(a) results in an induced charge current in the contact along the $z$-direction.

To estimate the size of this effect, it's important to account for the attenuation of the spin current which occurs at the perovskite-contact interface.  This attenuation, known as spin memory loss, results from interface disorder (present over an effective interface thickness $t_I$) and the reduced spin diffusion length at the interface (denoted $\ell_I^{\rm sf}$) due to the disorder \cite{eid2002absence}.  Spin diffusion models with spin memory loss yield the following expression for the 2-dimensional charge current $J_c$ induced in the contact due to the inverse spin Hall effect \cite{rojas2014spin}:
\begin{eqnarray}
J_c = \theta_{sh} Q^{yx} \ell_N^{\rm sf} \left(\frac{r_{sI}}{r_{sI}\cosh\left(\delta\right)+ r_{sN}\sinh\left(\delta\right)}\right) \label{eq:ishe}
\end{eqnarray}
where $\delta = t_I/\ell_I^{\rm sf}$, $\ell_N^{\rm sf}$ is the spin diffusion length in the contact, $r_{sI}=r_b/\delta$, with $r_b$ the interfacial resistance, and $r_{sN}=\rho\times  \ell_N^{\rm sf}$, with $\rho$ the bulk resistivity of the contact.  The term in parenthesis in Eq. \ref{eq:ishe} represents the suppression of transmitted spin current due to spin memory loss.  We estimate the interfacial resistance via the Sharvin relation: $r_b = \left(2e^2/h\times \pi k_F^2\right)^{-1}$, where $k_F$ is determined by the photogenerated carrier density.  We take a photon flux of $\phi_0=10^{25}~{\rm m^{-2}\cdot s^{-1}}$, $\rho=10^{-7}~{\rm \Omega\cdot m}$, and $\ell_N^{\rm sf}=3~{\rm nm}$.  The value of $\delta$ depends on the quality of the interface, which is difficult to anticipate.
Taking $\delta=2$, we find a spin memory loss attenuation of 0.25 and a resultant 2-d current density of $3\times 10^{-5}~{\rm A/m}$.  For a contact layer thickness of 10 nm and sample length of 4 mm in the $z$-direction, this corresponds to an induced voltage of approximately $0.1~{\rm \mu V}$.  We reiterate that the symmetry breaking direction of the crystal must be uniform over the sample length.  This should be possible given recently developed growth techniques which yield millimeter-sized single crystal grains \cite{shi2015low,nie2015high}.

\section{Summary} \label{sec:summary}

In summary, we report on the spin-dependent response of the organic-inorganic hybrid perovskite CH$_3$NH$_3$PbI$_3$ to unpolarized light.  We focused on the effect of Rashba spin-orbit coupling in the electronic structure, which leads to the generation of ballistic spin currents, and to the diffusive generation and transport of spin.  By considering a simple geometry, we provide analytic expressions which show the scaling of the spin-dependent response with material parameters, and can provide estimates for the photon flux required to generate measurable and potentially useful spin currents and densities.  We propose that the spin density can be used to infer the charge current distribution within the material, which would elucidate the role of grain boundaries in charge transport.  It is worth noting that the CH$_3$NH$_3$PbI$_3$ we investigated herein is just one of the large family of the organic-inorganic hybrid materials ABX$_3$ (A: CH$_3$NH$_3$, HC(NH$_2$)$_2$; B: Pb, Sn; X: Cl, Br, I). It is expected that other hybrid perovskites can also exhibit the spin responses we considered here, depending on the spin-orbit coupling and the degree of inversion symmetry breaking.

\acknowledgments{We acknowledge helpful conversations with Allan MacDonald, Mark Stiles, and Tom Silva.  J. L. acknowledges support under the Cooperative Research Agreement between the University of Maryland and the National Institute of Standards and Technology Center for Nanoscale Science and Technology, Award 70NANB10H193, through the University of Maryland.}

\appendix*
\section{Solution to Charge and Spin Diffusion Equations}

Here we discuss the solution to the charge and spin diffusion equations.  As described in the main text, we assume that the gradient of the ballistic spin current is of the form: $\tilde{\mu} \alpha\beta^2\tau_k \phi_0\exp\left(-\beta y\right)$.  The coupled charge-spin diffusion equations in 1 dimension (along the $y$-direction) are then:
\begin{widetext}
\begin{eqnarray}
-D\frac{\partial^2 n}{\partial y^2} + \frac{n}{\tau_n} + \Gamma_{sc} \frac{\partial S_x}{\partial y}  &=& \phi_0\beta  \exp\left(-\beta y\right)\label{eq:dd1a} \\
-D\frac{\partial^2 S_x}{\partial y^2} + \frac{S_x}{\tau_s} + \Gamma_{sc} \frac{\partial n}{\partial y}  &=&  -\tilde{\mu} \alpha\beta^2\tau_k\phi_0\exp\left(-\beta y\right) \label{eq:dd2a}
\end{eqnarray}

The general solution to Eqs. (\ref{eq:dd1a}-\ref{eq:dd2a}) is:
\begin{eqnarray}
n\left(y\right) &=& \sum_{j=1}^5 c_j n_j \exp\left(i k_j y\right)  \\
s\left(y\right) &=& \sum_{j=1}^5 c_j s_j \exp\left(i k_j y\right),
\end{eqnarray}
where the wave vectors (decay constants) are given by
\begin{eqnarray}
k_{1,2}&=&\pm \sqrt{\frac{-1-\lambda^2-\Gamma^2 - f}{2 L_d^2}}\\
k_{3,4}&=&\pm \sqrt{\frac{-1-\lambda^2-\Gamma^2 + f}{2 L_d^2}}\\
k_5&=&i\beta,
\end{eqnarray}
where $\lambda^2 =\tau_n/\tau_s$, $\Gamma=\Gamma_{sc}\sqrt{\tau_n/D}$, $L_d=\sqrt{D \tau_n}$.  The charge and spin density basis functions are given by:

\begin{eqnarray}
n_{1,2} &=&\pm\frac{\phi_0 \beta\tau_n\left(1-\lambda^2+\Gamma^2+f \mp\alpha\tilde{\mu} \beta\tau_k \Gamma \sqrt{2\left(-1-\lambda^2-\Gamma^2-f\right)}\right)}{id_+}\\
n_{3,4} &=&\pm\frac{\phi_0\beta\tau_n\left(-1+\lambda^2-\Gamma^2+f \pm i \alpha\tilde{\mu} \beta\tau_k \Gamma \sqrt{2\left(-1-\lambda^2-\Gamma^2+f\right)}\right)}{id_-}\\
s_{1,2} &=& -\frac{\phi_0\beta\tau_n\left(\Gamma \sqrt{2\left(-1-\lambda^2-\Gamma^2-f\right)} \pm i \alpha\tilde{\mu} \beta\tau_k \left(-1+\lambda^2+\Gamma^2+f\right)\right)}{id_+}\\
s_{3,4} &=& +\frac{\phi_0\beta\tau_n\left(\Gamma \sqrt{2\left(-1-\lambda^2-\Gamma^2+f\right)} \pm i \alpha\tilde{\mu} \beta\tau_k \left(-1+\lambda^2+\Gamma^2- f\right)\right)}{id_-}\\
n_5&=&\frac{i\phi_0\beta\tau_n\left(\lambda^2-\beta^2\left(1+\alpha\tilde{\mu}\Gamma\right)\right)}{d_5}\\
s_5&=&\frac{i\phi_0\beta^2\tau_n\left(\Gamma+\alpha\tilde{\mu}\left(\beta^2-1\right)\right)}{d_5}
\end{eqnarray}

written in terms of the following defined quantities:
\begin{eqnarray}
f&=&\sqrt{-4\lambda^2+\left(1+\lambda^2+\Gamma^2\right)^2}\\
d_{\pm}&=&f \sqrt{8\left(-1-\lambda^2-\Gamma^2\mp f\right)}\\
d_5&=&\left(\lambda-\beta\right)\left(\lambda+\beta\right)\left(1-\beta^2\right)-\beta^2\Gamma^2
\end{eqnarray}
The boundary conditions for the problem described in the main text are given below:
\begin{eqnarray}
\frac{\partial n(L)}{\partial y} &=& 0 \\
\frac{\partial S_x(L)}{\partial y} &=& 0 \\
n(0)&=&0 \\
S_x(0)&=&0
\end{eqnarray}
These 4 equations, together with $c_5=1$, fix the coefficients $c_j$ of the general solution.

For realistic values of the parameters, the spin lifetime is very short compared to the charge lifetime, and the resulting spin densities are small compared to the charge densities.  For this reason, \eref{eq:dd1a} can be easily solved by neglecting the spin source term.  The solution for the spin density is then found by solving \eref{eq:dd2a}, using the approximate charge density source term.
\end{widetext}

\bibliographystyle{apsrev}
\bibliography{ref}

\begin{thebibliography}{74}
\bibitem[{foo({\natexlab{a}})}]{footnote}
\bibinfo{note}{As discussed in Ref. 26, the diffusion equations follow from
  evaluting the spin-charge diffusion propagator. Going from 2 to 3 dimensions
  requires an additional integration over wave vectors in the additional
  dimension. This modifies the numerical prefactor of the constants which
  appear in the diffusion equation, but does not modify the general form of the
  diffusion equation.}
\bibitem[{foo({\natexlab{b}})}]{footnote2}
\bibinfo{note}{The Kerr signal is proportional to the spin density and the
  Faraday angle of the material. GaAs has an unusually large Faraday angle, so
  the spin densities needed for a measurable signal in CH$_3$NH$_3$PbI$_3$ may
  be larger than that of GaAs.}
\end{thebibliography}

\end{document}